\documentclass[twocolumn,showpacs,preprintnumbers,amsmath,amssymb]{revtex4}

\usepackage{graphicx}
\usepackage{bm}

\begin{document}


\title{Entangling neutrons via successive scattering from a substrate}


\author{M. Avellino, S. Bose, A. J. Fisher}
\affiliation{Department of Physics and Astronomy and London Centre for Nanotechnology,\\University College London,\\Gower Street, London WC1E
6BT}


\date{\today}

\begin{abstract}
This letter details a simple scheme to entangle two neutrons by successive scattering from a macroscopic sample. In zero magnetic field the entanglement falls as the sample size increases. However, by applying a field and tuning the momentum of the neutrons, one can achieve a substantial degree of entanglement irrespective of the size of the sample.
\end{abstract}

\pacs{113.5}

\maketitle


Neutrons are ideal model fermions for the study of quantum effects. They are easily initialized to a known state, have well-defined degrees of freedom and a small interaction cross-section, which limits coupling to the environment. Neutrons can also exhibit EPR-type correlations between distinct degrees of freedom, but so far these have only been observed for single particles, specifically between the spatial and spin parts of the neutron's wave function, and the neutron's energy \cite{rauch08}. In this letter, we put forward a scheme to create entangled states of two uncorrelated neutrons, hence opening up to further experimental study the fermionic analog of non-classical photon optics.

The aim of our work recalls recent studies on the use of a macroscopic sample
as both a mediator and an entanglement
reservoir, as well as a vast literature on scattering entanglement mediators and sequential generation of entangled states (\cite{dechiara05,cirac07,compagno04,yuasa05,schon05,yuasa06,costa06,ciccarello07} and references therein). However, as opposed to previous proposals, our scheme does not require single-state preparation, manipulation, ancillary measurements or pre-existing entanglement anywhere in the system: the entanglement between the neutrons arises purely as a consequence of their successive interaction with the sample.

We model our sample as a regular lattice of localized electrons, such as one might find in a ferromagnetic insulator. The sample is at zero temperature, and subject to a static magnetic field in the $\mathbf{\hat{z}}$ direction. We neglect the nuclei, and assume the electrons interact via a translationally invariant potential which conserves the total spin quantum number $S_z$, such as a ferromagnetic Heisenberg exchange interaction. Disregarding boundary effects, the free hamiltonian of the sample then reads:
\begin{equation}\label{h0}
\small H_0=-J\sum_{\langle ij \rangle}\bm{\sigma^i \cdot \sigma^j}+B_z\sum_{i=1}^N\mathbf{\sigma_z^i}
\end{equation}
with $J>0$ and $B_z>0$, where $\langle ij \rangle$ indicates the sum over nearest-neighbouring pairs, \textit{J} is the magnitude of the exchange
coupling constant, $B_z$ is the strength of the field, \textit{N} is the number of spins in the sample,
$\bm{\sigma}^i=\left(\sigma_x^i,\sigma_y^i,\sigma_z^i\right)$,
and the $\sigma_{\alpha}^i$ are the Pauli spin matrices for
spin \textit{i}. We use the ket $|0\rangle$ ($|1\rangle$) to denote the spin-up (down)
eigenstate of $\sigma_z$, and
indicate with $|j\rangle$ a state in which all the spins in the sample
except that at site \textit{j} are in state $|0\rangle$.

The protocol begins by initializing the sample to a known pure state, chosen such that the overall state of the neutrons and the sample contains at most two spin flips. We work with sample state $|\psi_i^A\rangle=\frac{1}{\sqrt{N}}\sum_{j=1}^N|j\rangle$, which is a uniform superposition of $|j\rangle$-states and thus contains a single spin flip. Similar results can be obtained by choosing $|\psi_i^B\rangle=|000..0\rangle$ provided one adjusts the neutron polarization accordingly. Hence, we do not rely on $|\psi_i\rangle$ being entangled, which distinguishes our protocol from an entanglement extraction scheme (cfr. \cite{dechiara05}). Both initial states can be prepared by applying an electron paramagnetic resonance pulse to the sample once it has relaxed to its ground state. In the absence of dissipative processes, the states $|\psi_i^{A,B}\rangle$ are stable, as they are eigenstates of $H_0$. In practice, their lifetimes will be limited by the classical spin relaxation time $T_1$ of the system. However, the success of the protocol will not be affected provided the entangling part of the scheme is completed within time $T_1$.

After a certain period $\tau_f\ll T_1$ of free evolution under
the effect of (\ref{h0}), the sample is irradiated with a beam of ultra-cold neutrons (UCNs) with momentum $k_z\mathbf{\hat{z}}$, prepared in $|0\rangle$ for state $|\psi_i^A\rangle$, or $\alpha|0\rangle+\beta|1\rangle$ for state $|\psi_i^B\rangle$, with $\alpha=\beta=\frac{1}{\sqrt{2}}$ in zero field, and $\alpha=0$, $\beta=1$ above a threshold field $B_t$, defined below. We assume the
intensity of the source is low so only one
neutron at a time scatters from the sample, with an arbitrary time delay between scatterings. This is reasonable because neutrons are weakly interacting particles, so for a typical flux the probability of many neutrons scattering at once is small. We model the scattering events as finite-time interactions of the neutrons with a composite hamiltonian of the form:
\begin{equation}\label{hgen}
\small \mathcal{H}_m=V_0 +\lambda\sum_{l}\bm{\sigma_n}^m
\cdot\lbrack \mathbf{\hat{Q}}\times \left(\bm{\sigma}_l\times
\mathbf{\hat{Q}}\right)\rbrack \:e^{i \mathbf{Q}\cdot\bm{R}_l},
\end{equation}
where the first term is a spin-independent potential extending over the finite volume $D^3$ of the sample, and the second, spin-dependent term arises from the magnetic dipole interaction between the neutrons and the sample \cite{fisher93}. Here, $\lambda$ is a coupling strength, $\bm{\sigma_n}^m$ is the spin of
neutron \textit{m}, $\mathbf{\hat{Q}}$ is the direction of the
neutron scattering wavevector
$\mathbf{Q}=\mathbf{k_i}-\mathbf{k_f}$, and $\bm{\sigma}_l$ is
the spin of the electron at position $\mathbf{R}_l$ in the sample. The value of $\lambda$ is determined by $\lambda=-g_n\mu_n g_e\mu_B\mu_0D^{-3}$, where $g_n$ is the neutron \textit{g}-factor, $\mu_n$ is the nuclear magneton, $g_e$ is the electron \textit{g}-factor, $D^3=Na_0^3$ and $a_0$ is the lattice constant. For the results shown, we define our energy and time units such that $\hbar=\mu_0=\mu_B=m_e=1$, and set $B_t=0.1\lambda N$.

\begin{figure}[h]
\resizebox{5cm}{!}{\includegraphics{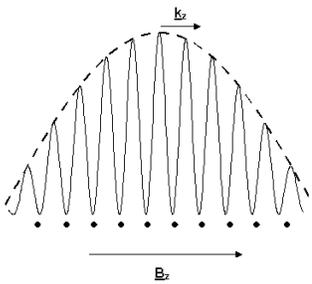}}
\caption{\label{scatt geom}\footnotesize{A one-dimensional snapshot of the scattering geometry at some fixed moment $t>\tau_f$. The sample sees the neutron as a wavepacket, the peak of which travels with momentum $k_z \mathbf{\hat{z}}$.} }
\end{figure}

We assume the sample `sees' the neutron as a de-localized wavepacket rather than a localized particle. As a result, the neutron can be thought to interact with all the spins in the sample simultaneously, which further distinguishes our scheme from an entanglement extraction protocol (FIG. \ref{scatt geom}). For this assumption to hold, the incident and scattered neutron waves must be in phase over the whole sample, hence the neutron coherence volume must be of order $D^3$. To maximize the scattered neutron flux we choose \textbf{Q} to be a reciprocal lattice vector, so that the phase factor in equation (\ref{hgen}) reduces to unity.

Let us detect neutrons scattered in the forward direction, for which $\mathbf{Q}=0$. To account for the rapid variation of $\mathbf{\hat{Q}}$ in the vicinity of $\mathbf{Q}=0$, we replace the double cross product in equation (\ref{hgen}) with its average value over a small sphere of radius $\epsilon \ll R^{-1}_l$, centred on $\mathbf{Q}=0$. One finds this average to be proportional to $\bm{\sigma}_l$. If we absorb this proportionality constant into the value of $\lambda$, the hamiltonian $\mathcal{H}_m$ can be approximated by the following expression:
\begin{equation}\label{h1 ex}
\small \mathcal{H}_m=V_0+\lambda\:\left(\sigma_{xm}
\cdot\sum_{i=1}^N \sigma_{x}^i+\sigma_{ym} \cdot\sum_{i=1}^N
\sigma_{y}^i+\sigma_{zm}
\cdot\sum_{i=1}^N \sigma_{z}^i\right),
\end{equation}
where $\sigma_{\alpha m}$ is the $\alpha$-component of the spin of the interacting neutron, which we label with \textit{m}, and the identity operation on the non-interacting neutron is understood. This hamiltonian gives rise to an exchange-type coupling between the neutron and the sample.

Let us suppose the first neutron arrives at the sample at time $t=\tau_f$ and remains coupled to it for a finite time $\tau$. At time $t=\tau_f+\tau$ it departs, and
the sample undergoes a second period of free evolution $\tau_f^\prime$, which is followed by the scattering of the second neutron. We assume the durations of the scattering events are equal, so that the second neutron also interacts with the sample for a time $\tau$. This is realistic, as $\tau$ is determined by the scattering process. The sequential nature of the interactions is assured by the fact that $\lbrack\mathcal{H}_1,\mathcal{H}_2\rbrack\neq0$. This condition must hold if any entanglement is to be produced \cite{brennen02}.

The initial state of the system as a whole can be written as $|\psi_0\rangle=|0\rangle_2|0\rangle_1|\psi_i\rangle$,
where the subscripts refer to the neutron indices and $|\psi_i\rangle$ is defined above. As $|\psi_0\rangle$ is pure, the scattered state $|\psi_f\rangle$ can be obtained by straightforward time-evolution using the canonical operator $U(\mathcal{H},\tau)=\exp{\lbrack-i\mathcal{H}\tau}\rbrack$, such that $|\psi_f\rangle=U(\mathcal{H}_2,\tau)U(\mathcal{H}_1,\tau)|\psi_0\rangle$. This is a slightly atypical way of treating a scattering problem, usually solved within the \textit{S}-matrix formalism which removes any explicit time dependence \cite{gell-mann53}. However, one can show that the two methods agree to first order in $V_0$ and $\lambda$, provided $\tau=D|k_z|^{-1}$. This condition supplies us with a physical interpretation of the quantum-mechanical time parameter: in the present context, it is the classical time a neutron of momentum $k_z\mathbf{\hat{z}}$ would take to travel a distance \textit{D}. This can be tuned simply by adjusting the neutron momentum. We commit a full analysis of this result to a future publication.

\begin{figure}[h]
\resizebox{5.5cm}{!}{\includegraphics{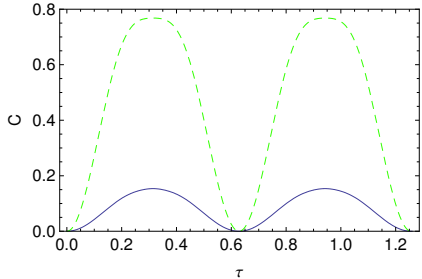}}
\caption{\label{four spins}The evolution of the concurrence as a function of the interaction time $\tau$ for $N=4$ and $B_z=0$. The green dashed curve relates to initial state $|\psi_i^A\rangle$ while the blue solid curve relates to initial state $|\psi_i^B\rangle$.}
\end{figure}

We quantify the entanglement between the neutrons using the concurrence as defined by Wooters
\textit{et al.} \cite{wooters00}, and assume throughout this paper that $J=\frac{1}{4}$. The choice of $J$ is entirely arbitrary, as it has no bearing on the behaviour of the concurrence. For illustrative purposes we set $\lambda=1$. Changing this value would simply re-scale our energy and time units.
\begin{widetext}
Through some algebra, it can be shown that the concurrence between the neutrons for initial state $|\psi_i^A\rangle$ is given by the following expression (we found no closed form for $|\psi_i^B\rangle$):
\begin{equation}\label{conc}
\small C(\lambda,N,B_z,\tau)=\frac{8\sqrt{2}N\lambda^2\sin^2{\tilde{\phi}\tau}}{\tilde{\phi}^{3}\lbrack B_z+\lambda(1-N)+\tilde{\phi}\rbrack}\sqrt{\lbrack\tilde{\phi}^{2}+\tilde{\phi}(\lambda-\lambda N+B_z)-2\lambda^2N\rbrack\lbrack\tilde{\phi}^2-4\lambda^2 N\sin^2{\tilde{\phi}\tau}\rbrack},
\end{equation}
\end{widetext}
with:
\begin{equation}
\small\tilde{\phi}\equiv\phi(\lambda,N,B_z)=\sqrt{B_z^2 - 2 B_z \lambda \left(N-1\right) + \lambda^2(N+1)^2}.
\end{equation}
In zero field, the concurrence shows regular oscillations as a function of $\tau$ (FIG. \ref{four spins}). The
period of these oscillations is determined by the energy splitting of the eigenstates of $\mathcal{H}_m$ which correspond to the spin-flip being shared between the interacting neutron and the sample. For initial state $|\psi_i^A\rangle$ these oscillations persist at finite fields, but for $|\psi_i^B\rangle$ the behaviour of the concurrence is more complex.

\begin{figure}[h]
\resizebox{7cm}{!}{\includegraphics{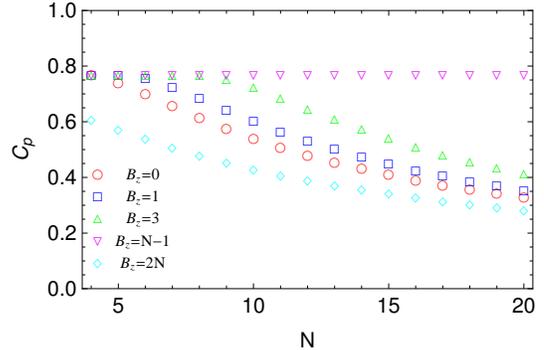}}
\caption{\label{co fn b}(Colour online) The dependence of the peak concurrence $\mathcal{C}_p$ on the magnitude of the field $B_z$ for initial state $|\psi_i^A\rangle$, with $\lambda=1$. Qualitatively similar results are found for initial state $|\psi_i^B\rangle$.}
\end{figure}

At $B_z=0$, the peak value of the concurrence $\mathcal{C}_p$ falls roughly as $N^{-1}$. However, this effect can be countered by switching on the field. For all \textit{N}, $\mathcal{C}_p$ is improved by the field, provided $B_z$ does not exceed a rough upper limit of $2\lambda N$. By maximizing Equation (\ref{conc}) with respect to $B_z$, one can show that the optimal field strength is in fact $B_z^*=\lambda(N-1)$ (FIG. \ref{co fn b}). Equation (\ref{conc}) then reads:
\begin{equation}\label{cred}
\small C(N,\lambda(N-1),\tau)=2|\cos{(2\lambda\sqrt{N}\tau)}|\sin^2{(2\lambda\sqrt{N}\tau)}.
\end{equation}
Hence, we calculate the optimal interaction time:
\begin{equation}\label{taustar}
\small \tau^*=\frac{1}{4\lambda\sqrt{N}}\cos^{-1}\left({-\frac{1}{3}}\right).
\end{equation}
For all \textit{N}, the maximum concurrence is then $\mathcal{C}_p=0.77$ (FIG. \ref{co fn b}). The corresponding state of the neutrons has the form $|\psi_f^{A,n}\rangle = \mu|00\rangle+\nu|01\rangle+\xi|10\rangle$, with $|\mu|^2\approx0.11$, $|\nu|^2\approx0.67$, $|\xi|^2\approx0.22$, $\frac{\nu}{\mu}\approx\sqrt{6}\exp^{i\frac{8\pi}{9}}$ and $\frac{\xi}{\mu}=\sqrt{2}\exp^{-i\frac{\pi}{2}}$.

The behaviour of the concurrence is independent of the period of free evolution between scatterings. This property may not extend to all entanglement measures, but is essential for the purpose of experimental implementation because we have no way of tuning $\tau_f^\prime$. In zero field the reason for the invariance is clear, as $H_0$ can be written as a multiple of the identity matrix \textbf{I}. In non-zero field, the interpretation remains outstanding. Quantitatively, the invariance under $\tau_f^\prime$ seems specific to the class of interaction
Hamiltonians $\lbrace\mathcal{H}_2\rbrace$ which conserve $S_z$. Hence, we infer that $U(\mathcal{H}_2,\tau)$ acts as a `decoding' operation, which extracts the signature deposited in the sample by the first neutron. It follows that for the neutrons to become correlated the second neutron must scatter before this `signature' disappears, i.e. within the spin coherence time $T_2$ of the material.

To verify the neutrons have become entangled we use a witness operator \textit{W} \cite{horodecki96,terhal00}. It has been shown that if the state $|\psi_f^n\rangle$ has some overlap with a `target' state of the form $|\varphi\rangle=\alpha|01\rangle+\beta|10\rangle$, there exists an optimal witness $W_{opt}$ which can be decomposed as follows \cite{guhne02}:
\begin{widetext}
\begin{equation}\label{witness}
\small W_{opt}=\alpha^2|z^+z^+\rangle\langle z^+z^+|+\beta^2|z^-z^-\rangle\langle z^-z^-|
+\alpha\beta\left(|x^+x^+\rangle\langle x^+x^+|+|x^-x^-\rangle\langle x^-x^-|-|y^+y^-\rangle\langle y^+y^-|-|y^-y^+\rangle\langle y^-y^+|\right),
\end{equation}
\end{widetext}
where $|x^{\pm}\rangle$, $|y^{\pm}\rangle$ and $|z^{\pm}\rangle$ are the spin-up and down eigenstates of the Pauli matrices $\sigma_x$, $\sigma_y$ and $\sigma_z$, respectively. Such a witness could be measured with as few as three device settings provided one could detect all outgoing neutrons and measure each component of their spin. This might be achieved by performing a Stern-Gerlach experiment in an arbitrary direction, which is conceptually possible though challenging from a technical viewpoint.

Finally, we examine the major sources of experimental uncertainty, such as errors in calibrating the magnetic field or the interaction time. Let us set a lower limit of $\mathcal{C}_p=0.7$ for the peak concurrence, and assume we operate either at optimal field or at optimal time. The allowed spread in $\tau$ and $B_z$ can then be approximated by the relations $\Delta \tau\approx10^{-3}\lambda^{-1}N^{-\frac{1}{2}}$ and $\Delta B_z\approx\lambda\sqrt{N}$. The fractional uncertainty in $\tau$ is therefore independent of \textit{N}, whereas the fractional uncertainty in $B_z$ is roughly proportional to $N^{-\frac{1}{2}}$. These relations yield stringent but not unsurmountable experimental requirements, given the precision to which neutron velocities and static magnetic fields can be calibrated \cite{aynajian08,li01,baciak03}.

Our proposal has an optical analogue in previous work by Haroche \textit{et al.} \cite{haroche97} on entangling pairs of atoms by exchange of a single photon in a high-Q cavity. However, we underline two important differences. Firstly, we assume both neutrons are prepared in the same state, contrary to the requirement in \cite{haroche97} that the first atom be excited and the second be in its ground state. Secondly, we assume the interaction time for both neutrons is the same. These alterations render our proposal a realistic solid-state analogue of \cite{haroche97}, as it is currently impossible to prepare two successive neutrons in different spin states and with different momenta.

So far, we have worked in natural units. Returning now to SI units, we use the technical specifications of the PF2 source of UCNs at the Institut Laue-Langevin in Grenoble to gauge some of the experimental requirements of our scheme \cite{steyerl86}. First, we estimate the required spin-relaxation time by imposing that $T_1$ be greater than the time taken by the neutrons to reach the sample. For UCNs with velocity $v=7$ ms$^{-1}$ and a flight path of $10^{-2}-1$ m this might require $T_1\approx 10^{-2}-1$ s, which is achievable in materials such as phosphorus-doped silicon or N@C$_{60}$ \cite{tyrishkin03,spaeth96}. Second, we require that the phase coherence time $T_2$ be greater than the time between scatterings. For a neutron flux $F=10^8$ m$^{-2}$s$^{-1}$ and a sample area of order $10^{-2}$ m$^{2}$, we find $T_2\approx 1$ $\mu$s, also attainable at low temperature \cite{tyrishkin03}.

Next, we require the neutron coherence volume to be comparable to the size of the sample. This condition yields an uncertainty relation between coherence length and momentum along a certain direction $\alpha$, such that $\Delta p_{\alpha}\approx \hbar\Delta L_{\alpha}^{-1}$. Assuming our sample were, say, 10 cm long, enforcing this condition would require the neutron velocity to be exact to one part in $10^6$, which is challenging but perhaps not unrealistic given recent progress in neutron spin-echo spectroscopy \cite{bayrakci06}.

Finally, we address the structural properties of the sample and the robustness with respect to experimental uncertainties. Given the form of $\lambda$, $B_z^*$ and $\tau^*$, one finds $B_z^*\approx10^{-32}(\frac{a_0}{\mathrm{m}})^{-3}$ T, $\tau^*\approx10^{21}(\frac{a_0}{\mathrm{m}})^{-3}N^{\frac{1}{2}}$ s and $D=v\tau^*$. Assuming an optimal field of $10^{-2}$ T and a sample 10 cm long, these relations yield $a_0=10^{-10}$ m and $N=10^{14}$, which are both attainable values. The allowed spread in the magnetic field and the neutron velocity is then $(\Delta B_z/B_z^*)\approx10^{-9}$ and $(\Delta v/v)\approx10^{-1}$. On both counts, this level of precision is within the capabilities of current experimental apparatus \cite{aynajian08,li01,baciak03}.

In conclusion, we have presented a simple scheme to create measurable entanglement between uncorrelated neutrons. An experimental realization would certainly be challenging, owing to the difficulty of detecting forward-scattered neutrons and performing arbitrary measurements on their spin. However, given the speed of progress in the field, such an experiment is perhaps not far beyond the reach of current neutron scattering facilities.

This research is part of QIP IRC www.qipirc.org (GR/S82176/01). The authors thank Christian Ruegg, Des McMorrow, Steve Bramwell, Tom Fennel and Francesco Ciccarello for many insightful discussions and suggestions. SB thanks EPSRC for an Advanced Research
Fellowship and the Royal Society and the Wolfson Foundation.

\end{document}